# Investigation and Automating Extraction of Thumbnails Produced by Image viewers

Wybren van der Meer
National Police of the Netherlands

The Netherlands
wybren.van-der-meer@ucdconnect.ie

Kim-Kwang Raymond Choo,
The University of Texas at San Antonio
USA
raymond.choo@fulbrightmail.org

Nhien-An Le-Khac, M-Tahar Kechadi
School of Computer Science
University College Dublin,
Belfield, Dublin 4, Ireland
{an.lekhac,tahar.kechadi}@ucd.ie

*Abstract*— Today, in digital forensics, images normally provide important information within an investigation. However, not all images may still be available within a forensic digital investigation as they were all deleted for example. Data carving can be used in this case to retrieve deleted images but the carving time is normally significant and these images can be moreover overwritten by other data. One of the solutions is to look at thumbnails of images that are no longer available. These thumbnails can often be found within databases created by either operating systems or image viewers. In literature, most research and practical focus on the extraction of thumbnails from databases created by the operating system. There is a little research working on the thumbnails created by the image reviewers as these thumbnails are application-driven in terms of pre-defined sizes, adjustments and storage location. Eventually, thumbnail databases from image viewers are significant forensic artefacts for investigators as these programs deal with large amounts of images. However, investigating these databases so far is still manual or semi-automatic task that leads to the huge amount of forensic time. Therefore, in this paper we propose a new approach of automating extraction of thumbnails produced by image viewers. We also test our approach with popular image viewers in different storage structures and locations to show its robustness.

## I. Introduction

Internet has been a major and increasing concern for many years. Criminals are using numerous methods to access data in the highly lucrative Cybercrime business. Organized crime, as well as individual users, are benefiting from the internet to carry out illegal activity such as money laundering, drug dealing and sharing child abusive material. The new connected way of life has drawbacks. One of them is that it has become increasingly easier to acquire child abusive material (CAM) [1]. After acquiring the CAM, crimes can look at these materials by using image viewers. Standard image viewers integrated in operating system do not offer effectively organising and categorising of image collections. To effectively organize image collections, crimes normally turn to specialised image viewers such as Xnview [2] and ACDsee [3].

To effectively organise and categorise the images, the image viewers normally save information about the images (so-called image meta-data) [4]. The image meta-data may contain thumbnails, dates, EXIF information and other sorts of information that are valuable forensic artefacts. The image meta-data is saved inside databases [5] which reside on the hard disk of the machine which has the image viewer installed on it. These databases can also contain meta-data information about images, which are no longer available. These images can be deleted, or reside on other sources that are not available to access at that time. For example, images are stored in an encrypted volume [6] which is not mounted. If the image resides on an encrypted volume, the image may not be available for the investigator. Eventually, when the source image is not available, the databases of image meta-data from image viewers are the only important sources of information available to investigators.

The image meta-data we look at in this paper is the thumbnails. In the literature, most research and practical focus on the extraction of thumbnails from databases created by the operating system. There is a little research working on the thumbnails created by the image reviewers due to many challenges. The first challenge is the storage locations for the databases containing the thumbnails are differ for each image viewer [7] and there are moreover many image viewers in the market [8]. When a database containing thumbnails is located, another challenge is to extract available relevant information. These databases often have their own way of storing thumbnails. Consequently, traditional ways of extracting information could not be applied. An example of a traditional ways of extracting information is carving for images [9]. Carving tools take sector size of a hard disk into account [10]. When a carving tool finds an image, the tool often skips a certain amount of disk space. The skipping of a certain amount of disk space is done because partitions on storage devices have a dedicated minimal storage space for each file. Hence, the skipping of this minimal storage space of each file is done to make the carving process faster. Thumbnails within a database can however, ignore these minimal storage spaces, as the database itself is one file. A carving program does not know there are multiple thumbnails within one file and automatically skips the minimal storage space, a space where many other thumbnails may reside.

Currently available literature suggests that investigating image databases is still a fairly unknown practice, as not a lot of information was found about investigating image databases. Most information found was tied to current carving methods, databases and thumbnails in general. Besides, large number of thumbnails stored in databases is also a challenges in terms of the investigating time. Therefore, in this paper, we propose a new approach to locate and extract thumbnails from image viewers. First, this approach is to locate the image databases by examining the image viewers and then to search for changes made on the hard drive. The changes gave clues about were the image databases might be stored. When the image database is found, we then extract the thumbnails inside by using different techniques described in this paper because of three different storage structures observed. Another contribution of this paper is to automate the entire proposed process of investigation, using freely available tools and code to reduce the time needed for the forensic investigation.

The rest of this paper is organized as follow: Section II shows the related work of this research. We present the problem statement of investigating the thumbnails from image viewers in Section III. We describe our approach in Section IV. We show our experiments on different image viewers analysis results found in Section V. Finally, we conclude and discuss on future work in Section VI.

## II. RELATED WORK

In this section, a literature survey was conducted towards databases and carving. Subjects related to the investigation of image databases are also included.

To speed up the viewing of images the thumbnails has been proposed for a while [11]. Today, the size of images and photo's is more and more bigger as image sensors become more advanced. Thumbnails are therefore still needed today to speed things up, and to keep data usage low. This is also the case on local residing storage within a computer. When a computer only has to load the smaller thumbnails, the user will experience less waiting time. Fast storage like Solid State Drives is becoming cheaper and more common. It is however still cheaper to store large quantities, like photo collections, on traditional hard drives. At the moment of writing, Solid State Hard drives cost around €0.28 per GB [12], traditional hard drives cost around €0.028 per GB [13]. The reason why Solid State Drives are faster than the traditional spinning hard drives is that the Solid State Drives have no moving parts [14]. This makes the seek times [15] very short, and the read and write speed a lot higher

Thumbnails do not only offer a faster user experience, they also have other uses. Thumbnails are often created before a picture has been manipulated. The original picture can then be altered. The thumbnail associated to an image would then be different compared to the new altered image [16]. In this way the thumbnail can be used to check if the image has been altered. Thumbnails also offer a useful source of information about an image if the original is not present at all. When the original is lost, the investigator still has a smaller representation of an image to investigate.

One good way of investigating an image can be to reverse search the image. In this manner the source of the original image can be tracked down [17]. The reverse searching can also be done with just the thumbnail.

A paper has been written about a forensic analysis of Windows *thumbcache* files [18]. This paper deals with how thumbnails created by Windows are stored for different Windows versions. The versions Windows: 95, 98, ME, 2000, XP, 2003, Vista, 7 and 8 were analysed on how the thumbnails created by the operating system were stored. The paper proceeds with investigating how the Windows thumbnails are stored and how to extract them. A prototype tool for automated extraction of Windows thumbnails is presented. The prototype tool also takes information from the Windows index files, a feature which comparable tools seem to lack. The paper gives a good blueprint on how to design an automated tool for investigated thumbnails. It does however only focus on the thumbnails created by the different Windows operating systems. Thumbnails made by individual programs were not investigated.

In [19], authors focus on when Windows thumbnails are produced. In this paper different tests are conducted to see when and how Windows produces thumbnails. The paper concludes that sometimes thumbnails are even produced when the user does not browse related pictures. The investigator should always validate by other means if the user actually saw the pictures related to the found thumbnails. The actually viewing of a picture is a parameter and has been taken into account during the determining of the method to investigate thumbnails generated by image viewers.

Thumbnails produced by Android were also previously investigated [20]. In this paper authors focus on extracting thumbnails and the file information of the original files from the main *imgcache* file found within an Android installation. The *imgcache* stores the thumbnails of images encountered by the operating system itself. According to the authors, the filenames and other file information are also stored within this *imgcache* file. The authors were successful in extracting this information and finding out how the information was stored. Also the EXIF information is stored within the found thumbnails. Conducted tests revealed that thumbnails remained in the *imgcache* file, even after the original files were deleted. In order to find this information, authors used tools like scalpel [21] (carving tool), HxD [22] (Hex editor) and EXIF tool [23] (tool for finding EXIF information). This research shows that not only the thumbnail can be stored by a database, but that the EXIF information might also be extracted.

Many techniques can be used to create thumbnails [24]. There is also a wide choice of available software to create thumbnails, one them being Easy Thumbnail for Python. This tool is a good example of how many techniques there are to create thumbnails [25]. Examples of different techniques are scanning for borders using entropy, different options for output colours and different quality settings. In [26] these different techniques for creating thumbnails are taken into account when matching found thumbnails with the original images. The technique used for the matching takes six steps for making a thumbnail into account. These steps are cropping, pre-filter, down-sampling, post-filter, contrast and brightness adjustment, and JPEG compression. In the paper the importance of using thumbnails for authentication is emphasized. Using thumbnails for these goals is an important use for thumbnails in a forensic investigation. This underlines how important it is to find the thumbnails. The authors downloaded a number of pictures to test their methods. This is a good way to validate findings.

Databases are an important source of information to conduct a forensic investigation [27] [28] [29]. An example of a widely used database is the SQLite database. This database is used a lot in mobile operating systems present on smartphones [30]. The storing of information is so flexible within a SQLite database, often all the information a smartphone app needs is stored within a SQLite database. In [31], the main database used for thumbnails in OS X Mountain Lion was investigated. This database is a SQLite database called "*index.sqlite*". In the paper test were conducted to see which actions led to changes in this "*index.sqlite*" database. In the conducted research it became clear that for each browsed image, two rows were filled in the SQLite database. One row contained the information about the original file such as filename, size and folder. The other row contained information about the thumbnail that was created. Combining the different rows gave full information about the original file and the thumbnail. If entries remained after the original file was deleted, was not investigated. Findings were checked by browsing pictures on a new account on a newly installed machine. This check also led to the conclusion that the database was not removed when a user was removed from the system. The fact a database is able to store different kinds of data and retrieve this information makes it an ideal choice to store thumbnails, and information about the original pictures. The information about an image and the information of it can then be requested by a program with a single query.

Carving techniques focus on the recovery from raw data block available on a hard disk. There are different ways to carve for files. A basic method to carve for files is signature based carving [32]. Signature based carving looks for a header and a footer of a file. The header and footer are file signatures, which mark the beginning and end of a file. The signature based carving takes the data between the header and footer and makes it available as a new file. One of the most well-known file signatures is FFD8 (header) and FFD9 (footer), these mark the header and footer of a jpg image file. Signature based carving will only be able to successfully complete the recovering of files if they are not fragmented. For the recovery of fragmented files, more advanced techniques are needed. A more advanced technique is object validation. Object validation will try to open the file to check if this generates an error. This is however not always a valid method, as a stream of false data does not always induce an error [33].

Authors in [34] describe an effort is made to bring the different fragmented parts of a carved image together. With this method it would be possible to carve complete images, even if the image is stored fragmented. Authors used a technique which checks for different patterns in JPG compression. If the pattern in the end of one fragment matched the pattern in the beginning of one other fragment, statistical tests were conducted in order to validate if the fragments should be stitched together. This technique is not used by standard carving tools. Therefore, standard carving tools often do not carve fragmented images fully. The researched have brought the different techniques together in a product called MyKarve. The fragmentation of images is something that should be reckoned with in the own research.

Recovering data from databases may not be possible with carving, as the structure may be different. It is however possible to retrieve data from databases by other means. In the study of [35], some work has been done on recovering data from DBMS structures. Another subject to consider is compressed data within databases. There are methods for extracting the compressed streams, however, errors may occur [36].

III. PROBLEM STATEMENT

Images and their metadata can be a useful source of information within an investigation. The image itself may contain visual clues. The metadata may contain clues about the time and date when certain actions occurred. In order to use these clues, the images and their metadata must first be found. Investigation can be conducted well when the images are available on unencrypted data storage, and can be accessed and viewed. When an image is deleted, fewer options remain to investigate an image. Carving is a method, which can retrieve deleted files within a file system.

Carving is however not always successful, as files are sometimes returned fragmented. Also, carving also can't be done on encrypted data when the decryption key is unknown.

When the original images are no longer available, and cannot be retrieved by carving files, one has to look for different sources of information. One source may be forensic artefacts created when an image is viewed. An image can be browsed and viewed by the file explorer of an operating system. This may lead to the creation of a thumbnail. Also metadata of the file may be stored. The retrieving of thumbnails and metadata from artefacts within the file explorer of an operating system has been researched and will remain a subject with each new version of operating systems. Information is however not stored within the database of the operating system if the image was viewed by other means.

When an image is viewed by a specialized image viewer, forensics artefacts may be created by the viewer itself. The method used for storing the information by a viewer may differ between different brands, and can change with each new version. The viewer, like the operating system, can store information within a database. The storing of the information in a database has different purposes. One main consideration for storing thumbnails is the speed increase for showing many pictures within a gallery mode.

Each viewer could use a different storage method or database. There are also many image viewers available. The different storage methods for each found database, means different methods must be devised on how to extract the information needed. One method for extracting thumbnails is carving for files.

Carving techniques focuses on recovering files from partitions on a hard disk. Data within a database on the hard disk may have a different data structures then files residing in a partition on a hard disk. For this reason traditional forensic processes can have problems recovering all the thumbnails from the database. To be able to recover information from the databases containing images, the structure of these databases must be investigated. When the structure is found the information needed can be extracted. Without a specialized tool for this, the extraction of information and thumbnails remains a manual task for the digital forensic investigator. A manual task which needs many different tools, and many different steps. Considering the steps needed, investigating the databases produced by image viewers can be a time consuming task. Meaning the databases may not always be investigated, due to the time and effort needed for it. This can lead to useful information not being found. Streamlining the process of extracting the thumbnails and other information would mean the investigation would take less time. Taking less time, the investigation may be done more often. This could lead to useful information found more often.

Automating the different steps needed within the investigation would be a step towards streamlining the process. For this to be possible the different steps within the investigation must first be investigated. Answering the following research questions can do this:

• How are the images stored within the databases for popular image viewing software?

• What methods can be applied to identify database structures containing images?

• If there are images present, what is the best way to extract all of these images, without skipping any?

• How does a traditional forensic tool like Encase deal with these databases?

With the answers of these questions clear, effort can be made to write an automated script, which will save time on investigating information stored by image viewers.

## IV. ADOPTED APPROACH

There are two main stages in our proposed approach. First, it is the identification of storage location. This means we identify how different image viewers store the images and their thumbnails. The second stage is dealing with database structure containing images. Indeed, we have developed an automate method to identify the structure of the databases containing images, and to extract the identified images from the databases. We describe these stages in details as follows.

To investigate the storage methods of the image viewers, we apply the following steps: (i) Install different image viewers; (ii) Using image viewer to open and browse images stored in external device and monitoring these activities.

By having the images on an external device, and by only browsing the images with the image viewers, effort was made to reduce the traces by other means than the image viewers. By accessing images from an external drive, effort was made to come as close to real life usage of such an image viewer as possible. Besides, the FolderChangesView tool [37] is used to log the changes to the hard drive. The monitoring of changes during the installing of the viewers, and during the viewing of the images gave information about which files were written to.

By first investigating the files that were changed most often, effort was made to find the databases. Also the location of the files that had changes was a factor that was looked at. Combining these factors lead to files that had the most change of being a database. Monitoring the file changes on the live system had the advantage of being able to look at the results in a real time fashion. In this way the file changes occurring exactly during the image browsing could be more easily determined.

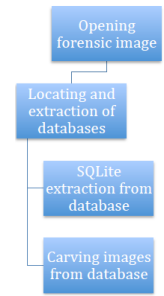

Fig. 1. Workflow of proposed approach

After the first stage, suspect files are extracted. Next, in the second stage, we examine the suspect files to locate images stored. By looking at the header and other plain text contents of the file, sometimes the structure of the database could be determined. If the type of file could not be determined this way, the file was carved to see if any of the viewed images returned. If the suspected file is a database and contains the pictures as raw data, this process would have carved the pictures out, making the thumbnails available for further investigation. If the header of the file is a SQLite database, the file will be opened in a SQLite database viewer. A SQLite database viewer explores the fields and tables contained within the SQLite database file. Such a field within the database can contain binary data. This is then called a data "blob". If such a field contains the binary data of an image, the data can be copied out and written to a new file, which then can be opened like a normal image file.

When the structure of the databases containing images, and the methods to extract the images, were already known, a method is determined on how to extract information from the databases containing images in an automated way. Our approach is to write a script, which was able to automate the extraction of found thumbnails. In other words, the script has to be able to find known databases containing images, and be able to extract said images. There are four phases in our automate process (Figure 1): (i) opening the forensic image; (ii) locating and extraction the databases; (iii) parsing SQLite databases and (iv) carving databases for images.

## V. EXPERIMENTS AND DISCUSSION

In our experiments, we tested our approach implemented as a Python script with seven popular image viewers listed as Xnview, Irfanview, Faststone, Adobe Lightroom, Zoner photo studio, Acdsee 19 and Google Photos. We describe first of all how these image viewers store images and how we can retrieve images from them by using our automate approach mentioned above. We are using the following datasets to test our approach:

- Images from www.pexels.com (open stock photo site, 4862 images, no EXIF).
- NEC animal toy dataset (5000 images from 60 toy animals, no EXIF)
- Describable Textures Dataset (5500 images of textures, no EXIF)
- ImageCLEF2011 collection (18000 images, EXIF, not embedded)[39]
- Imagenet Large Scale Visual Recognition Challenge 2015 test dataset (11142 images, no EXIF)
- California-ND Q0mex 2013 dataset (704 images, EXIF embedded)[40]
- Self-shot collection (392 images, EXIF embedded)

We also used a traditional forensic tool like Encase to investigate the forensic image to compare with our approach as using Encase would be able to find the databases containing images. We also aim to check if Encase is also able to extract the thumbnails from the databases. During this testing, an estimation was made on how the entire process of extracting thumbnails from image databases could be automated within Encase.

A. Tools and storage methods

*1) Xnview:*

After browsing the prepared storage device with images, numerous changes were made to the file XnView.db located in the folder path

C:\Users\Administrator\AppData\Roaming\XnView\. The XnView.db is a SQlite database that stores metadata about the stored thumbnails. The thumbnails itself were stored in the "Datas" table. The data of the thumbnails seemed obfuscated. This was confirmed by the fact that standard carving tools were unable to retrieve images from the database.

*2) Irfanview:*

The tool Irfanview is an image viewer, which at the first glance seemed to be an image viewer to only show single files. However, under the file option, a thumbnail option is visible. After selecting the thumbnail option, a folder tree of the machine becomes visible. When selecting a map under this thumbnail viewer, the tool will start to collect thumbnails of each picture available in the folder. During the scanning of the folder and the building of thumbnails, no files were written which could be related to Irfanview. It was possible the thumbnails were only loaded into memory. To validate this a folder with large amount of pictures was opened and the memory usage was monitored. The process "i_view32.exe" was using an increasing amount of memory as the thumbnails were created. A screenshot was made at the end of the scanning of the images and is shown in the image below (Figure 2).

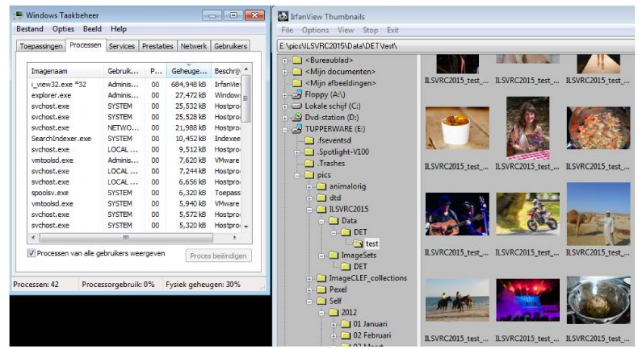

Fig. 2. Irfanview thumbnails

*3) Faststone:*

*Faststone* is a freeware image viewer tool. This tool also showed a browsable folder tree of the machine. The software was started and images were browsed. The file "FSViewer.db" located at the folder path

C:\Users\Administrator\AppData\Roaming\FastStone\FSIV \ showed a high number of changes during the monitoring of file and folder changes. This file turned out to be a *TinyDB* database. This was done by looking at the data with the hex viewer "xxd". The thumbnails within were however not obfuscated and carving techniques were able to carve images out of the database. The tool scalpel was used for the carving process. The outcome of the scalpel tool is shown in the picture below (Figure 3).

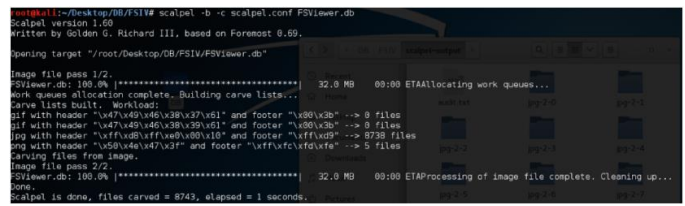

Fig. 3. Carving Fastone thumbnails

*4) Adobe Lightroom:*

Adobe Lightroom is a professional tool for managing and editing photos. It features many functions like adjusting colour, cutting photos and many more functions. To test this software a trial version of the software was downloaded and installed. The prepared images were browsed and the file changes were monitored. It became clear that the program created many files with the .db extension. These DB files turned out to be sqlite databases. Many of these database files seemed to log data about what has been done to images in the program. The database "root pixel.db" under the folder

"C:\Users\Administrator\Pictures\Lightroom\Lightroom CatalogPreviews.lrdata" contained some thumbnails. The thumbnails were found under the Table "RootPixels" in the row "jpegData" of this SQlite database. The data of this field was not encoded and could be extracted to recreate a viewable picture, as shown below (Figure 4).

Carving the data of the SQlite database with a standard tool did not yield any images.

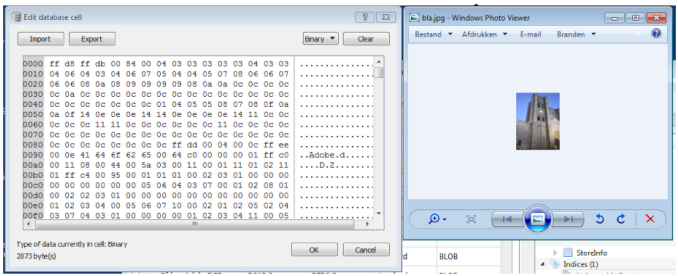

Fig. 4. Carving Adobe Lightroom thumbnails

*5) Zoner photo studio:*

Zoner photo studio is a tool for managing and editing photos. It features functions like adjusting colour, cropping and many more functions needed for a professional photographer. In this tool the folder structure was visible in a window on the left side. When browsing through this folder tree the images of the selected folder became visible on the screen. During the monitoring of folder and file changes one file stood out due to the large number of changes. The file that stood out was "data.zoner-index-cache" residing in the folder

"C:\Users\Administrator\AppData\Local\Zoner\ZPS 18\ZPSCache.dat". This file was identified as an SQlite database and contained the metadata related to the files viewed within the photo studio. In this case metadata such as created date, filepath of the picture, author of the picture file and many more was saved within the database. The database did not contain the thumbnails for the viewed images. Thumbnails were saved as image files, although missing their extension in the filename, under the path "C:\Users\Administrator\AppData\Local\Zoner\ZPS18\ZPSCache.dat\000\". Using Kali Linux, a preview of this folder has been made and is shown in the picture below (Figure 5).

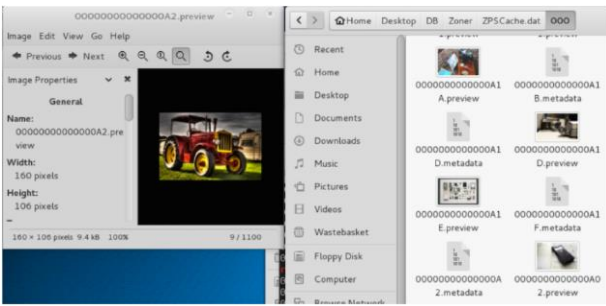

Fig. 5. Zoner thumbnails

*6) ACDsee 19:*

ACDsee is a tool for organising images. To test this tool a trial version was used. The tool notices the user of the existence of the database as soon an external media was browsed, due to the fact the tool asks the user if data of the external media should be stored in the database. After browsing the prepared images on the USBdrive, due to the logging of folder changes, it became clear that ACDsee stored its data under the path

"C:\Users\Administrator\AppData\Local\ACDSystems\Catalogs\190\Default\". Within this folder, tree files stood out due to the number of writes to the files, size and the filenames; thumb1.fpt, thumb2.ftp and thumb3.ftp. These file where examined using kali Linux. To get a quick view of how many jpg file headers were found the command "xxd <filename> | egrep 'ffd8|ff d8' | wc –l " was used of the thumb.ftp files. An overview of the results is listed below.

TABLE I.    THUMBNAIL FOUND

| Filename | Number of hits |
|---|---|
| Thumb1.ftp | 1406 |
| Thumb2.ftp | 1342 |
| Thumb3.ftp | 1269 |

The Thumb.ftp files were also carved using the command "recoverjpeg –b 1 <filename>". The option –b1 stands for the blocksize of 1. This means the tool will ignore the standard block size of 512 and will recover images if they are written to a file continuously. In the image below the results of this process are shown (Figure 6):

Fig. 6. ACDsee thumbnails

Upon visual inspection the recover files seemed to be the same except they were differing in size, with the Thumb1.ftp database containing the files with the largest size and best quality.

*7) Google Photos:*

Google Picasa is a popular tool for managing images and photos. However, as of March 2016, Google will stop its support for this tool and refers to Google Photos for future use. In order to test if Google Photos leaves any traces, the service was tested with prepared images. During the testing it became clear that the service of Google Photos was only useable within a web browser. There was an installable tool available for the uploading of pictures. This tool did however not view or save any pictures that were uploaded. Due to the fact Google Photos is only available via a web browser, there was not a local database that saved thumbnails. The locations that were written to during the testing were linked to the web browser Firefox that was used for the testing.

*8) General discussion*

Having investigated the way popular image viewers storing their data shows that different storage methods were used: (i) SQLite database; (ii) Custom database; (iii) Thumbnails as normal files in file system and (iv) in memory. The storage methods seem to differ so much from each other that one method of retrieving the information is not sufficient to retrieve the information of all the tools. For this reason multiple methods were used for retrieving the information from popular image viewers. In some cases the image viewers did not store any data about the viewed images on the hard disk. These tools stored the data in the memory. This leads to missing information when only the forensic image is investigated. Using live forensics, it may be possible to retrieve information from these tools that do not store information on the hard disk. Live forensics was however not in the scope of this paper. The reason is that live forensics cannot be conducted on a forensic image, but have to be conducted on the live running machine.

The investigated tools, which only stored data in memory, were Irfanview and Google photos. The database used by XnView was a SQLite database. The tables within the SQLite database, which likely contained the image data, seemed obfuscated. Another tool using a SQLite database for storing information about viewed images was Adobe Lightroom. The database was investigated with a SQLite database viewer. Such a viewer provides an easy way to examine the different fields stored within the database. Adobe Lightroom is a tool with many options and methods for viewing images.

The tool Zoner Photo Studio also uses a SQLite database. This database stored the metadata of the viewed pictures. The thumbnails were stored as files within the file system. The way the thumbnails were stored by Zoner Photo Studio is a good example of a different way of storing them, in comparison to the other tools. The different method of storage also meant a different method of extracting the images should be used in comparison to when the images were stored within a database.

Faststone and ACDsee 19 are tools that proved to have a custom database for storing the thumbnails of the viewed images. These databases did however not use compression or encryption to store the thumbnails. Not using encryption or compression means the data within is available in raw format, which means carving for files will retrieve the files contained in the database. With these tools, carving the databases for images retrieved the thumbnails of the images viewed within the virtual machine. It is possible more information was contained within these databases, and could have been extracted using tools for accessing exactly the used databases. These tools were however unavailable during the research, which is why only carving methods were used.

*9) Discussion on the automate approach:*
To find the databases the decision was made to search for the filenames of the databases within the file systems present in the forensic image. The downside of searching for filenames is when the creators of the image viewing tool change the name of the database, this has to be changed in the tool also. However this method saves time opposed to extracting all possible databases found on the image, and then having to determine which the right one is. Furthermore the searching within the file system is a less time consuming method then to carve the entire forensic image for database files. If a file matching the name of the database was found, it was extracted for further processing.

To extract the images from the SQlite database, the exact table containing the images was extracted. By writing the data of these fields to a new jpg file, the images become available to the users of the script. The extracting of the fields containing the image data was done using the sqlite3 module within Python. There was also the idea to extract all SQlite databases available on the image, and then to parse each entry of all SQlite databases. This could however greatly increase the time the script needs to run as the data that needs to be processed increases. Also this would not only extract the images found in databases of image viewers, but also images which were found elsewhere. Such a module would be more suited for a script, which has as goal to extract as many images from a forensic image as possible. This may be a goal for a future research.

For the carving of images from the databases containing the images in a raw format, an own function was created. The reason an own function was created for the carving is that a module that already did this with the options needed for the databases, was not found. The found modules within Python that did carving were focused on carving from a file system. To carve the files form the database file, the database file was first converted to hex values. Regular expressions were used to look for headers and footers of jpg images in the hex values of the database files. To keep this regular expression simple the choice was made to only extract jpg images. Making a regular expression search for many different headers and footers would also increase the time the script would need to carve trough the databases. Using many headers and footers may also return many false positive files.

*B. Encase*

As mentioned previously, we also did experiments with Encase to check if it is also able to extract the thumbnails from the databases.

*1) Carving:*
The main carving option within Encase is able to carve within files. This option must be checked at the start of the investigation, and is only able carving within the allocated space, or to carve within all files. The options are shown in the image below. Encase was able to carve the images from the databases that had raw image data within them. However, this process took a lot of time, as it also did the carving process for all other files in the image.

*2) SQLite:*
The SQLite databases are not dealt with through Encase itself. Encase can however send the selected file to another program. In this case the Encase was able to send the file to a SQLite viewer. Within the SQLite viewer the images were extracted

*3) Discussion:*
Using the forensic Encase tool, the databases containing images were investigated. All the actions in Encase used to retrieve the images were manual. Some steps may be automated using Enscripts. The parsing of SQLite databases needed to be done outside of Encase, as Encase cannot process these files alone. Encase sending the SQLite databases to an external tool makes it more complicated to automate the entire process of extracting images from the databases.

The carving process in Encase seems to carve all the files within the image. This is a time consuming action to do because the tools has to scan for all files in allocated space of the hard drive investigated. The carving of the entire image will return many images available on the hard disk, images such as icons. Analysing all images may cost a lot of time, compared to when only the viewed images from image viewers are extracted.

## VI. CONCLUSION AND FUTURE WORK

The time it takes to extract thumbnails from custom databases used by programs is one of the problems stated in this paper. This could lead to the investigation of thumbnails being less common. An attempt has been made to tackle this problem within this paper, by making a script which can extract the thumbnails in an automated way. First the storage methods of the thumbnails were investigated. By using different tools it became apparent that the following storage methods were used: the images were stored in SQLite database, in other database in raw form, in raw form as one file per thumbnail, or only in the memory. The automated script dealt with this by looking for the thumbnails in the databases that were stored on the hard disk, not in the memory. During the investigation of the storage methods the file names and locations of the database became apparent. This information was used to identify and extract the databases in the script. To extract the thumbnails from the found databases different techniques were used. With the goal of reducing the time needed to extract thumbnails from custom databases used by image viewing programs in mind, the script was designed to work as efficient as possible. The script took around one minute to complete on the forensic image used in this paper, returned thousands of thumbnails.

Locating and extracting the thumbnails was also tested with the use of the forensic software Encase. During this investigation it seemed not all steps could be automated. This was due to the fact a new case had to be made and the carving options to be set. The carving options were so that these could return the thumbnails. The downside however was that then all files were scanned, taking large amounts of time. Also SQLite databases were not dealt with in the program itself. Automating the process using Encase may be done in a future work.